\documentstyle[preprint,aps]{revtex}
\widetext
\draft
\tighten

\begin{document}

\title{Principle of Mach, the Equivalence Principle and  concepts of
inertial mass} \author{Andrew E. Chubykalo}
 \maketitle
  \begin{center}
 {\it Escuela de F\'{\i}sica, Universidad Aut\'onoma de Zacatacas\\ 98068
Zacatecas, ZAC., M\'EXICO}
%\date{October 24, 1995}
\end{center}
$$ $$ \baselineskip 7mm
\begin{abstract}
 A study of kinematics of a 2-body
 system is used to show that the Mach principle,  previously rejected by
general relativity, can still serve as an alternative to the concept of
absolute space, if one takes into account that the background of distant
stars (galaxies) determines {\it both} the inertial and the gravitational
masses of a body.
\end{abstract} $$ $$ $$ $$ {\bf PACS}:
04.20.-q,01.55.+b \clearpage

It is well  known that in classical mechanics there were two distinct
concepts of inertial mass and absolute space: that of Newton ({\it in
which inertial mass is a property of a body with respect to absolute
space}), and  that of Mach ({\it in which inertial mass is the property
determined by the masses of  distant stars}) [1]. It is also known that
when construction general  relativity, Einstein started with the Mach
principle, but had to reject  it thereafter (e.g.,see [2]) because of its
unagreement with the Equivalence  principle. However there still is no
general agreement among scientists  about the necessity of total rejection
of the Mach principle (e.g., see [3-7]).

Let there be a  two-body system with the inertial and gravitational masses
$m_{\tt i},M_{\tt i}$  and $m_{\tt g},M_{\tt g}$, accordingly. Let it be
surrounded by some  collection of distant stars, at rest with respect to
the center of inertia  (CI) of the system. Let these two bodies rotate
around their common CI  with some angular velocity $\omega$. To be more
specific, let us assume  that the distance between the CI of the system
and the nearest star is  much greater than the distance $R$ between the
bodies, which, in its turn, is much  greater than the size of the bodies;
this will allow us to consider them as  material points. Let us also
assume that their velocities are much  smaller than the speed of light,
which will allow us to use the equations  of classical mechanics. We shall
further assume that the angular velocity of  rotation $\omega$ is such that
the bodies can get closer to each other  only because of the loss of
energy due to gravitational radiation, i.e.,  in the absence of such
radiation, the distance between the two bodies  would remain constant.

Let us analyze the situation arising in the hypothetical case in which the
distant stars {\it disappear}, assuming that at that instant the
bodies rotated. In such setup, there are only {\it two} logically possible
situations:  \bigskip

a){\it  either the kinematics of the relative motion of the system
changes, i.e., an observer located on one of the bodies perceives a
picture different from the one he would see in the presence of fixed
distant stars;} \bigskip

b) {\it or the observer does not perceive any difference in the kinematic
pattern of the relative motion of the bodies.}
$$$$ The situation (a) can
happen only if the Mach principle holds in the form in which it has been
known so far [1], since in Newton description by definition there is no
change  in the state of the system under a removal of sufficiently distant
stars (we  set up that the {\it gravitational} influence of this stars is
infinitesimal).  A change in the state of the system would be possible
only due to a change  in the relationship between the inertial and
gravitational masses  (by virtue of the Mach principle, the absence of
distant stars must lead  to a strong decrease in the values of inertial
masses, and, therefore,  to sharp increase in the relative acceleration of
their mutual approach).  Such a situation would be contradict the
principle of equality of  inertial ($m_{\tt i}$) and gravitational
 ($m_{\tt g}$) masses but, generally speaking, we cannot consider the {\it
 principle of equivalence} like one of {\it underlying axioms} of general
 relativity.\footnote{See, e.g., brilliant work of M. Sachs ``On the Logical
 Status of Equivalence Principles in General Relativity Theory" [8]. That
 is, there is no $\alpha\ \pi\varrho\iota o\varrho\iota$ reason why it
 should necessarily follow that $m_{\tt i}=m_{\tt g}$, even though
 experimental observations confirm this equality to high accuracy.} The
 case (a) was studied already rather well (profound analysis of this case
 is given by P. Graneau [9] and A. Assis [10,11]). Following these works we
 have to infer that if the case (a) happens we can {\it either} say that
 the gravitational constant $\gamma$ {\it or} the inertial mass $m_{\tt
 i}$ of the test body will be a function of the amount and distribution of
 distant bodies (stars and galaxies).

 However, we cannot test neither the case (a) nor the case (b) (although
 a direct check-up of the Mach principle can be realized in the
 laboratory, the effect will be too small to be detected). Therefore, we
 still {\it must} consider the case (b) and case (a) like enjoying equal
  rights. In this paper we shall prove that if would be realized the case
  (b) the Mach principle can still remain an alternative  to the Newton
  concept of absolute space, on the one hand, and  allows for the equality
  of the inertial and gravitational masses, on  the other hand (at least,
  in classical mechanics).

  For detailed analysis of the case (b) we are going to consider two
  hypothetical situations: distant bodies (the rest of stars of the
  Universe) {\it do not exist} and the Universe only {\it consists} of two
  bodies $m$ and $M$; distant stars (galaxies) {\it exist}.

In the case (b) both  concepts of inertial mass could be valid - the
Newton and {\it modified}\,\footnote{Generally accepted Mach principle can
be realized in the case (a) only (see above and [9-11])} Mach ones (see
below).  Notice that in this reasoning we have to assume that aside from
``local action" (in the Faraday terminology) and independently of it there
is ``action-at-a-distance" ({\it instantaneous action}) in nature.

So, in order to modify the Mach principle and still keep the kinematic
equivalence of the two concepts of space for  a circular motion, one has
to assume that distant stars determine {\it both} inertial y gravitational
properties of a body\footnote{Below we shall prove  that this assumption
has got rather strict arguments, at least, in classical mechanics (see
 {\bf Theorem})}. We shall call such a concept ``quasi-Mach". Notice that
 in  quasi-Mach case of situation (b), the masses cannot be equal to zero,
 because  each body serves as a background for the second one.

Notice that these bodies may turn around their common CI, following either
elliptic or circular orbits. In our Gedanken experiment we chose the
 circular orbits. Such a selection of circular orbits may seem unfounded
 at  the first sight, but in fact it can be easily explained: in order to
 obtain  the relationships in which we are interested (see below), we have
 to choose such  kind of motion whose relative kinematics does not depend
 on whether it is  Newton or the quasi-Mach concept is true. In the case
 of elliptic motion,  if the stars disappear, the observer will see
 either $(i)$  ``oscillations" of bodies with respect to each other ({\it
 aphelion-perihelion}),  which would automatically signify the validity of
 Newton's concept,  or $(ii)$ these ``oscillations" will cease, which
 would mean that there  is an influence of stars on the masses  of the
 bodies.  Thus in both  cases $(i)$ and$(ii)$ we would obtain a {\it
 unique} answer in favor  of one of the two concepts. In reality such an
 experiment is naturally  impossible. However, in the case of circular
 orbits there is no unique  choice of a valid concept for the observer,
 which will allow us to  retain the assumption that the two concept are
 equally justified.

If we choose as the ``true"  concept the quasi-Mach one, then by equating
kinematic properties of the  same type in the {\it presence} and in the
{\it absence} of distant stars,  we can obtain a relationship between the
``old" (in the {\it presence} of the rest of the
matter) and the ``new" (in its {\it absence},  accordingly) masses.

We  sall use the equations of classical mechanics for the ``new" masses,
 while for the ``old" masses we shall use the equation for gravitational
 radiation of a  system of two bodies rotating around their common CI.

In the case of  ``new" masses, we consider two bodies in an absolutely
empty space, which come closer to each other under the influence of the
gravitational force. Remember that speaking about a rotation of such a
system has no meaning any more since both the stars and the notion of an
absolute space are absent in this concept, so that the only ``real"
coordinate here is the  {\it distance} between the two bodies. The second
law of Newton and the inverse-square law lead to a relative acceleration
in the two-body system:  \begin{equation} \ddot{x}=\gamma\frac{M_{\tt g}
m_{\tt g}}{M_{\tt i} m_{\tt i}}\biggl(\frac{M_{\tt i}+m_{\tt
i}}{x^2}\biggr), \end{equation} where $x$ is the distance between the
bodies; $M_{\tt i},m_{\tt i}$ and $M_{\tt g},m_{\tt g}$ are the inertial
and gravitational masses of the bodies $M$ and $m$, accordingly.

In the case of  ``old" masses, the two bodies rotate around their common
CI in the presence of stars (or in the absolute space, which in this case
is the same). The potential energy of the system has the form $$
\varepsilon_{\tt pot}=\gamma\frac{M_{\tt g} m_{\tt g}}{r},
$$
where $r$ is the distance between bodies. From the condition of equality
forces and the rotational frequencies, we can obtain the expressions for
the linear velocities of the bodies:  $$ V^2_{\tt M}=\gamma\frac{m_{\tt
g}M_{\tt g}m_{\tt i}} {M_{\tt i}(M_{\tt i}+m_{\tt i})r}; \qquad V^2_{\tt
m}=\gamma\frac{m_{\tt g}M_{\tt g}M_{\tt i}} {m_{\tt i}(M_{\tt i}+m_{\tt
i})r}.  $$ Substituting them into the equations for the kinetic energy of
the bodies, we find $$ \varepsilon^{\tt tot}_{\tt k}=\gamma\frac{M_{\tt
g}m_{\tt g}} {2r}, $$ where $\varepsilon^{\tt tot}_{\tt k}$ is the total
kinetic energy of the system (remember that we consider these bodies as
material points).  Notice that for a circular motion, the total kinetic
energy of bodies  depends {\it only on their gravitational masses}, rather
than inertial ones.  This fact allows us to prove following {\bf
theorem}: {\it In the  framework of} {\tt classical mechanics} {\it the
gravitational mass determines the ``inertia" of material body}.

{\tt Proof}: {\small So, let some body $m$ (with inertial mass  $m_i$) to
move along a straight line with constant velocity $V$. Its kinetic energy
is:  \begin{equation} K = \frac{m_i V^{2}}{2} \end{equation}

From kinematic point of view the movement with constant velocity along a
straight line and (with constant {\it linear} velocity) along a
circumference of infinity radius are equivalents.

We can consider also the movement of the similar body (with same inertial
mass $m_i$) as circumference movement around the other body $M$. Now we
can require that kinetic energy and {\it linear} velocity of body $m$ to
be equal to that of the case (2). Nothing can forbid us to do it. Now from
the equivalence force conditions we have:  \begin{equation} \frac{m_i
V^{2}}{R} = {\gamma}\frac{m_g M_g}{R^{2}} \end{equation} where $m_g,M_g$
are gravitational masses of bodies $m$ and $M$, $R$ is distance between
$m$ and $M$. Expressing "$V^{2}$" from (3) and substituting it into the
formula of the kinetic energy of the body $m$ obtain:  \begin{equation} K
= {\gamma}\frac{m_g M_g}{2R} \end{equation}

Let us now to increase $M_g$ and $R$ conserving the same time value of
$K$. In this case $M_g(R)$ and $R(M_g)$ are one-to-one functions. It is
obvious that if $K$, $m_g$ and ${\gamma}$ are constants, $M_g(R)$ and $R$
will be {\it linear dependent} functions, i.e.  $$ M_g(R) = C\cdot R, $$
where $C$ is some suitable dimensional constant. After tending $R$ to
infinity (conserving the same time values of $K,m_g$ and ${\gamma}$) we
obtain from (4) \begin{equation} K=m_g\frac{{\gamma} C}{2} \end{equation}
where "${\gamma} C$" has dimension of the "$V^{2}$". It means that (5) can
be rewritten as \begin{equation} K=m_g \beta\frac{V^{2}}{2} \end{equation}
here $\beta$ is non-defined constant.

Recalling above-mentioned notice (equivalence between straight line
movement and the same one along the circumference of the infinite radius),
we conclude that kinetic energy of the body moving along the straight line
with the constant velocity {\it is proportional to the gravitational
mass}. Comparing (6) and (2) we obtain the equivalence \begin{equation}
m_i=m_g \beta, \end{equation} where $\beta$ is a constant non-defined in
frames of the above-mentioned considerations.}

Now we can assumed that $\beta =1$ in (6) and as result, the total
mechanical energy of the system (see above) is \begin{equation}
\varepsilon=-\gamma\frac{Mm}{2r}.  \end{equation} Here and below we shall
skip the indices ``i" and ``g" according to the meaning of the problem.

The radiation rate of the gravitational energy during a circular motion
has the form [12]:  \begin{equation}
-\frac{d\varepsilon}{dt}=32\gamma\biggl(\frac
{mM}{m+M}\biggr)\frac{r^4\omega^6}{5c^2}.  \end{equation} From (8) we have
\begin{equation} \frac{d\varepsilon}{dt}=\gamma\frac{Mm}{2r^2}\frac
{dr}{dt}.  \end{equation} Then we obtain \begin{equation}
\omega^6=\gamma^3\frac{(m+M)^3}{r^9}.  \end{equation} Substituting (10)
and (11) into (9) and differentiating the resulting expression with
respect to time, we can find the relative acceleration of the mutual
approach of the bodies:  \begin{equation}
\frac{d^2r}{dt^2}=-3\gamma^6\biggl[\frac{64}{5c^5} mM(m+M)\biggr]^2r^{-7}.
\end{equation} Now, denoting in (1) both masses by the index ``n" and in
(12), by ``o" (``{\it new}" and ``{\it old}" masses), and comparing these
two expressions, we obtain the desired relationship \begin{equation}
M_{\tt n}+m_{\tt n}=\alpha\frac{ m^2_{\tt o}M^2_{\tt o} (m_{\tt o}+M_{\tt
o})^2}{R^5}.  \end{equation} Here we denoted all constants coefficients by
$\alpha$, while $R$ is the distance between the bodies.

Now let us consider another situation: distant stars (galaxies) exist
``again" (remember that according to Mach, it is irrelevant whether it is
the background or the bodies that rotate).  Then $m_{\tt n}$ becomes
$m_{\tt o}$, and $M_{\tt n}$ becomes $M_{\tt o}$.  That is, $m_{\tt n}$
and $M_{\tt n}$ get multiplied by some factors $A$ and $B$ $(Am_{\tt
n}=m_{\tt o};  BM_{\tt n}=M_{\tt o})$ which are functions of th masses:
$A=A(M_{\tt o},\Phi), B=B(m_{\tt o},\Phi)$, where $\Phi$ are masses of the
rest of stars. Then (13) can be written in the form \begin{equation}
M_{\tt n}+m_{\tt n}=\alpha\frac {m^2_{\tt n}M^2_{\tt n}A^2B^2(Am_{\tt
n}+BM_{\tt n})^2} {R^5}.  \end{equation} Using the fact that $\Phi\gg
m_{\tt o},M_{\tt o}$, we can expand $A(M_{\tt o},\Phi),B(m_{\tt o},\Phi)$
in a power series in small parameters.  In the zeroth approximation, $$
A\cong A(0,\Phi);\qquad B\cong B(0,\Phi).  $$ Since it is clear that the
functional form of $A$ and $B$ must be same, we can write \begin{equation}
A(0,\Phi)\equiv B(0,\Phi)=A(\Phi).  \end{equation} We do not possess more
information about the form of the function $A(\Phi)$, but we shall assume
that when $\Phi\rightarrow\infty$, $A\rightarrow const$, i.e., we shall
assume that $A$ is a constant specific for our Universe. Substituting (15)
into (14), we obtain \begin{equation} \alpha A^6 m^2_{\tt n}M^2_{\tt
n}(m_{\tt n}+M_{\tt n})=R^5.  \end{equation} The constant $A$ cannot be
determined within the framework of this problem. We have only shown the
necessity of its existence {\it if the quasi-Mach concept is true}. Now
let us note that there exists such $p$ that $M_{\tt n}=pm_{\tt n}$. The
Eq.(16) implies that \begin{equation} [p^2(1+p)]^{1/5}m_{\tt n}=(\alpha
A^6)^{-1/5}R.  \end{equation}

Thus we have shown that Newton solution of the problem of absolute space
and inertial mass, taking into account the requirements of general
relativity, is {\it non-unique} if the Mach principle is modified in a
suitable way. We have also shown that if we choose the quasi-Mach concept
as {\it true}, then in our approximation the masses of two bodies in an
empty Universe {\it are proportional to the distance between them} (17).
When the bodies approach each other, their masses tend to {\it zero} (we
mast not forget that our computation are based on the bodies being
pointlike, though this restriction is not a matter of principle). Notice
that if the masses of the bodies vanish when they come close to each
other, this would not signify that matter disappear, and therefore such
vanishing of the masses would mean that there is some conserved property
of matter in nature which, perhaps, is unrelated to the space-time
structure of the Universe. In the framework of the problem under
consideration we are unable to say anything more specific about this
property.  \bigskip

{\bf Acknowledgments}

I am grateful to Profs.  Jayant V. Narlikar, A. Assis and  V.
Dvoeglazov for many stimulating discussions and critical comments.

 \end{document}